\documentstyle[preprint,pra,aps,fleqn]{revtex}
\begin{document}
\title{{\bf Photon-added one-photon and two-photon nonlinear coherent states}}
\author{Xiaoguang Wang\thanks{%
email:xyw@aphy.iphy.ac.cn}}
\address{Laboratory of Optical Physics, Institute of Physics,Chinese Academy of
Sciences,\\
Beijing 100080, P.R.China \\ and  
CCAST(World Laboratory),P.O.Box 8730, Beijing 100080, P. R. China}
\date{\today}
\maketitle

\begin{abstract}
From the photon-added one-photon nonlinear coherent states $a^{\dagger m}|\alpha,f\rangle$, 
we introduce a new type
of nonlinear coherent states with negative
values of $m.$ The nonlinear coherent states corresponding to the positive
and negative values of $m$ are shown to be the result of nonunitarily
deforming the number states $|m\rangle $ and $|0\rangle $, respectively. As
an example, we study the sub-Poissonian statistics and squeezing effects of
the photon-added geometric states with negative values of $m$ in detail.
Finally we investigate the photon-added two-photon nonlinear coherent
states
and find they are still the two-photon nonlinear coherent states with
certain nonlinear functions.
\end{abstract}


\begin{center}
{\bf {\large 1. Introduction}}
\end{center}

Recently there has much interest in the study of nonlinear coherent
states(NLCSs)\cite{NLCS,NLCS1}, which are right-hand eigenstates of the
product of the boson annihilation operator $a$ and a nonlinear function $%
f(N) $ of the number operator $N$,

\begin{equation}
f(N)a|\alpha ,f\rangle =\alpha |\alpha ,f\rangle .
\end{equation}
Here $\alpha $ is a complex eigenvalue. It has been shown that a class of
NLCSs may appear as stationary states of the centre-of-mass motion of a
trapped ion\cite{NLCS}. These nonlinear coherent states exhibit nonclassical
features like squeezing and self-splitting.

Another type of interesting nonclassical states consists of the photon-added
states\cite{Agarwal,Dodonov}

\begin{equation}
|m,\psi \rangle =\frac{a^{\dagger m}|\psi \rangle }{\langle \psi
|a^ma^{\dagger m}|\psi \rangle },
\end{equation}
where $|\psi \rangle $ may be an arbitrary quantum state, $a^{\dagger }$ is
the boson creation operator, $m$ is a non-negative integer-the number of
added quanta. For the first time these states were introduced by Agarwal and
Tara\cite{Agarwal} as photon-added coherent states. The photon-added
squeezed states\cite{PASS} ,even(odd) photon-added states\cite{EOPAS} and
photon-added thermal state\cite{PATS} were also introduced and studied. The
photon-added states can be produced in the interaction of a two-level atom
with a cavity field initially prepared in the state $|\psi \rangle $\cite
{Agarwal}.

Sivakumar showed that the photon-added coherent states are nonlinear
coherent states\cite{Sivakumar}. As a generalization we showed a general
result that photon-added NLCSs(PANLCSs) are still NLCSs with different
nonlinear functions\cite{Wang1}. The PANLCSs are defined as

\begin{equation}
|m,\alpha ,f\rangle =\frac{a^{\dagger m}|\alpha ,f\rangle }{\langle \alpha
,f|a^ma^{\dagger m}|\alpha ,f\rangle }.
\end{equation}
They satisfy\cite{Wang1}

\begin{equation}
f(N-m)[1-m/(N+1)]a|m,\alpha ,f\rangle =\alpha |m,\alpha ,f\rangle .
\end{equation}
As seen from Eq.(4), the PANLCS is an NLCS with the nonlinear function $%
f(N-m)[1-m/(N+1)].$ Naturally Eq.(4) reduces to Eq.(1) when $m=0.$ The
well-known geometric states(GSs)\cite{GS} and negative binomial states(NBSs)%
\cite{NBS} are NLCSs\cite{Wang1}. Therefore, the photon-added GSs\cite
{Barnett,Wang2} and photon-added NBSs\cite{Wang3} are still NLCSs and are
special cases of the PANLCSs.

In the present paper we show that the PANLCSs are the result of nonunitarily
deforming the number state $|m\rangle .$ We introduce the PANLCS with
negative values of $m$ , which are the result of nonunitairily deforming the
vacuum state $|0\rangle $. As an example, we study the sub-Poissonian
statistics and squeezing effects of the photon-added geometric states with
negative values of $m$ in detail. We also investigate the photon-added 
two-photon nonlinear coherent states.

\begin{center}
{\bf {\large 2.The PANLCS as deformed number state $|m\rangle $}}
\end{center}

In this section we show that the PANLCS can be written as a nonunitarily
deformed number state. This is achieved by the method given by Shanta et al%
\cite{Shanta}. Here we give a brief review of the method.

Consider an annihilation operator $A$ which annihilates a set of number
states $|n_i\rangle ,i=1,2,...k.$ Then we can construct a sector $S_i$ by
repeatedly applying $A^{\dagger },$ the adjoint of $A,$ on the number state $%
|n_i\rangle .$ Thus we have $k$ sectors corresponding to the states that are
annihilated by $A.$ A given sector may turn out to be either of finite or
infinite dimension. If a sector, say $S_j,$ is of infinite dimension then
we can construct an operator $G_j^{\dagger }$ such that $[A,G_j^{\dagger
}]=1 $ holds in that sector. Then the eigenstates of $A$ can be written as $%
\exp (\alpha G_j^{\dagger })|n_j\rangle .$ If an operator $A$ is of the form 
$f(N)a^p,$ where $p$ is non-negative integer, such that it annihilates the
number state $|j\rangle $ then $G_j^{\dagger }$ is constructed as\cite
{Shanta}

\begin{equation}
G_j^{\dagger }=\frac 1pA^{\dagger }\frac 1{AA^{\dagger }}(a^{\dagger }a+p-j).
\end{equation}

It is interesting that the operator $f(N-m)[1-m/(N+1)]a$ in Eq.(4)
annihilates both the vacuum state $|0\rangle $ and Fock state 
%TCIMACRO{\TEXTsymbol{\vert}}
%BeginExpansion
\mbox{$\vert$}%
%EndExpansion
$m\rangle .$ The states between the vacuum state and Fock state 
%TCIMACRO{\TEXTsymbol{\vert}}
%BeginExpansion
\mbox{$\vert$}%
%EndExpansion
$m\rangle $ are not annihilated. To discuss the case of the PANLCS $%
|m,\alpha ,f\rangle $ let

\begin{equation}
A=f(N-m)[1-m/(N+1)]a,A^{\dagger }=a^{\dagger }f(N-m)[1-m/(N+1)]
\end{equation}
We construct sector $S_0$ by repeated applying $A^{\dagger }$ on the vacuum
state. $S_0$ is the set $|i\rangle ,i=0,1,2,...,m-1$ and it is of finite
dimension. The sector $S_m,$ built by the repeated application of $%
A^{\dagger }$ on $|m\rangle ,$ is the set $|i\rangle ,i=m,m+1,...$ and it is
of infinite dimension. Hence we can construct an operator $G^{\dagger }$
such that $[A,G_{}^{\dagger }]=1$ holds in $S_m.$ To construct $G^{\dagger },
$ we set $p=1$ and $j=m$ in Eq.(5) and this yields

\begin{equation}
G^{\dagger }=a^{\dagger }\frac 1{f(N-m)}
\end{equation}
In fact, by direct verification, we have

\begin{equation}
\lbrack f(N-m)[1-m/(N+1)]a,a^{\dagger }\frac 1{f(N-m)}]=1.
\end{equation}
Therefore the PANLCS can be written as

\begin{equation}
|m,\alpha ,f\rangle =\exp (G^{\dagger })|m\rangle =\exp [a^{\dagger }\frac 1{%
f(N-m)}]|m\rangle
\end{equation}
up to a normalization constant. From the above equation it is shown that the
PANLCS can be viewed as nonunitarily deformed Fock(number) state $|m\rangle
. $

\begin{center}
{\bf {\large 3.The PANLCS with negative $m$}}
\end{center}

The form of $A,$ given by Eq.(6), suggests that it is a well-defined
operator-valued function also for negative values of $m$ on the Fock space.
In this section the PANLCS with negative $m$ is constructed. Denoting the
the PANLCS with negtive $m$  by $|-m,\alpha ,f\rangle ,$ the equation to
determine them are

\begin{equation}
f(N+m)[1+m/(N+1)]a|-m,\alpha ,f\rangle =\alpha |-m,\alpha ,f\rangle .
\end{equation}
The operator $A=f(N+m)[1+m/(N+1)]a$ only annihilates the vacuum state. When $%
f(N)\equiv 1,$ the state $|-m,\alpha ,f\rangle $ reduces to that studied in
Ref.\cite{Sivakumar}. The sector $S_0,$ built by the repeated application of 
$A^{\dagger }=$ $a^{\dagger }f(N+m)[1+m/(N+1)]$ on $|0\rangle ,$ is the set $%
|i\rangle ,i=0,1,...$ and it is just the infinite dimensional Fock space. To
construct $G^{\dagger },$ corresponding to the operator $A=f(N+m)[1+m/(N+1)]a
$ , we set $p=1$ and $j=0$ in Eq.(5) and this yields

\begin{equation}
G^{\dagger }=a^{\dagger }\frac{N+1}{f(N+m)(N+m+1)}
\end{equation}
Thus the PANLCS with negative $m$ can be written as

\begin{equation}
|-m,\alpha ,f\rangle =\exp (G^{\dagger })|0\rangle =\exp [a^{\dagger }\frac{%
N+1}{f(N+m)(N+m+1)}]|0\rangle
\end{equation}
up to a normalization constant. The state $|-m,\alpha ,f\rangle $ is
obtained by nonunitarily deforming the vacuum state $|0\rangle $ while the
state $|m,\alpha ,f\rangle $ is obtained by nonunitarily deforming the Fock
state $|m\rangle .$

The PANLCS is obtained by the action of $a^{\dagger m}$ on the NLCS $|\alpha
,f\rangle $. The state $|-m,\alpha ,f\rangle $ can be written in a similar
form using the inverse operators $a^{-1}$ and $a^{\dagger -1}$.\cite{Mehta}
 These operators are defined in terms of their actions on the number state $%
|n\rangle $ as follows

\begin{eqnarray}
a^{-1}|n\rangle &=&\frac 1{\sqrt{n+1}}|n+1\rangle , \\
a^{\dagger -1}|n\rangle &=&\frac 1{\sqrt{n}}|n-1\rangle ,  \nonumber \\
a^{\dagger -1}|0\rangle &=&0.  \nonumber
\end{eqnarray}
Using these inverse operators the state $|-m,\alpha ,f\rangle $ can be
rewritten as
$|-m,\alpha ,f\rangle =a^{\dagger -m}a^{-m}|\alpha ,f^{\prime }\rangle$   
up to a normalization constant. Here $|\alpha ,f^{\prime }\rangle $ is the
NLCS with the nonlinear function $f^{\prime }(N)=f(N+m).$ The state $%
|-m,\alpha ,f\rangle $ is obtained by the action of the operator $a^{\dagger
-m}a^{-m}$ on the NLCS $|\alpha ,f^{\prime }\rangle $ while the state $%
|m,\alpha ,f\rangle $ is obtained by the action of the operator $a^{\dagger
m}$ on the NLCS $|\alpha ,f\rangle .$

From Eq.(12) the number state expansion of the PANLCS with negative $m$ can
be easily obtained as

\begin{equation}
|-m,\alpha ,f\rangle =\sum_{n=0}^\infty \frac{\alpha ^n\sqrt{n!}}{%
f(n+m-1)...f(0)(n+m)!}|n\rangle
\end{equation}
up to a normalization constant. The expansion is useful in the following
discussions.

\begin{center}
{\bf {\large 4.Photon-added geometric state with negative $m$}}
\end{center}

In this section we consider a special example of the PANLCS with negative $%
m, $ the photon-added geometric state with negative $m.$

The geometric state is defined as\cite{GS}

\begin{equation}
|\eta \rangle =\eta ^{1/2}\sum_{n=0}^\infty (1-\eta )^{n/2}|n\rangle ,\text{ 
}0<\eta <1,
\end{equation}
It satisfies

\begin{equation}
\frac 1{\sqrt{N+1}}a|\eta \rangle =\sqrt{1-\eta }|\eta \rangle .
\end{equation}
In comparison with Eq.(1), we see that the geometric state is an NLCS with
the nonlinear function $1/\sqrt{N+1}.$ The photon-added geometric state is
defined as

\begin{eqnarray}
|m,\eta \rangle &=&\frac{a^{\dagger m}|\eta \rangle }{\langle \eta
|a^ma^{\dagger m}|\eta \rangle } \\
&=&\eta ^{(m+1)/2}\sum_{n=0}^\infty {%
%TCIMACRO{\binom{m+n}n }
%BeginExpansion
{m+n \choose n}%
%EndExpansion
}^{n/2}(1-\eta )^{n/2}|n\rangle ,
\end{eqnarray}
which is just the negative binomial state introduced by Barnett\cite{Barnett}%
. We have studied the statistical properties and algebraic characteristics
of the photon-added geometric state in detail\cite{Wang2}. From Eqs.(4) and
(16) we get

\begin{equation}
\frac{\sqrt{N-m+1}}{N+1}a|m,\eta \rangle =\sqrt{1-\eta }|m,\eta \rangle .
\end{equation}

The state 
%TCIMACRO{\TEXTsymbol{\vert}}
%BeginExpansion
\mbox{$\vert$}%
%EndExpansion
$m,\eta \rangle $ is an NLCS with the nonlinear function $f(N)=\sqrt{N-m+1}%
/(N+1).$ When $m=0,$ Eq.(19) reduces to Eq.(16) as we expected.

We would like to study the state $|-m,\eta \rangle ,$ the photon-added
geometric state with negative values of $m,$ which satisfies

\begin{equation}
\frac{\sqrt{N+m+1}}{N+1}a|-m,\eta \rangle =\sqrt{1-\eta }|-m,\eta \rangle .
\end{equation}
From Eq.(14), the number state expansion of the state $|-m,\eta \rangle $ is
given by

\begin{equation}
|-m,\eta \rangle =\text{ }\sqrt{\frac{m!}{_2F_1(1,1;m+1;1-\eta )}}%
\sum_{n=0}^\infty (1-\eta )^{n/2}\sqrt{\frac{n!}{(n+m)!}}|n\rangle ,
\end{equation}
where $_2F_1(1,1;m+1;1-\eta )$ is the hypergeometric function.

The photon statistics of a quantum state can be conveniently studied by
Mandel's $Q$-parameter\cite{Q}

\begin{equation}
Q=\frac{\langle N^2\rangle -\langle N\rangle ^2-\langle N\rangle }{\langle
N\rangle }
\end{equation}
A negative $Q$ indicates that the photon number distribution is
sub-Poissonian and it is a nonclassical feature. A positive $Q$ indicates
the super-Poissonian distribution and $Q$=0 indicates Poissonian
distribution. The photon-added geometric state $|m,\eta \rangle $ can be
sub-Poissoian depending on the parameter $\eta $\cite{Wang2}. For the state $%
|-m,\eta \rangle ($Eq.(21)$),$ the mean value of $N^k$ is easily obtained as

\begin{equation}
\langle N^k\rangle =\frac{m!}{_2F_1(1,1;m+1;1-\eta )}\sum_{n=0}^\infty
n^k(1-\eta )^n\frac{n!}{(n+m)!}.
\end{equation}

In Fig.1 the $Q$-parameter, calculated using Eqs.(22) and (23), for the
state $|-m,\eta \rangle $ is shown as a function of $\eta .$ The $Q$%
-parameter is always greater than zero indicating that they are
super-Poissonian. For larger values of $\eta ,$ the $Q$-parameter is close
to zero since the state $|-m,\eta \rangle $ reduces to $|0\rangle $ in the
limit $\eta \rightarrow 1$.

\begin{center}
{\bf {\large 5.Squeezing in $|-m,\eta \rangle $}}
\end{center}

Define the quadrature operators $X$(coordinate) and $Y$ (momentum) by 
\begin{equation}
X=\frac 12(a+a^{\dagger }),Y=\frac 1{2i}(a-a^{\dagger }).
\end{equation}
Then their variances 
\begin{equation}
Var(X)=\langle X^2\rangle -\langle X\rangle ^2,Var(Y)=\langle Y^2\rangle
-\langle Y\rangle ^2
\end{equation}
obey the Heisenberg's uncertainty relation 
\begin{equation}
Var(X)Var(Y)\ge \frac 1{16}.
\end{equation}
If one of the variances is less than 1/4, the squeezing occurs. In the
present case, $\langle a\rangle $ and $\langle a^2\rangle $ are real. Thus,
the variances of $X$ and $Y$ can be written as 
\begin{eqnarray}
Var(X) &=&\frac 14+\frac 12(\langle a^{\dagger }a\rangle +\langle a^2\rangle
-2\langle a\rangle ^2), \\
Var(Y) &=&\frac 14+\frac 12(\langle a^{\dagger }a\rangle -\langle a^2\rangle
).
\end{eqnarray}

From Eq.(21), the expectation value $\langle -m,\eta |a^k|-m,\eta \rangle $
is directly obtained as

\begin{eqnarray}
\langle -m,\eta |a^k|-m,\eta \rangle  &=&\frac{m!}{_2F_1(1,1;m+1;1-\eta )} 
\nonumber \\
&&\sum_{n=0}^\infty (1-\eta )^{n+k/2}\frac{(n+k)!}{\sqrt{(n+m)!(n+m+k)!}}
\end{eqnarray}
The variances of $X$ and $Y$ can be calculated from Eqs.(27), (28) and (29).
In Fig.2 we show the variances of the quadrature operators $X$ and $Y$ as a
function of $\eta $ for different values of $m.$ The squeezing exists in the
quadrature $Y.$ For the quadrature $Y,$ the degree of the squeezing becomes
deep with the increase of the parameter $m.$ In the limit $\eta \rightarrow
1,$ the variances are all equal to $1/4.$ This is because the state $%
|-m,\eta \rangle $ reduces the vacuum state $|0\rangle $ in this limit.

\begin{center}
{\bf {\large 5. Photon-added two-photon nonlinear coherent states}}
\end{center}
In this section, we investigate the photon-added two-photon nonlinear
coherent states.
The two-photon nonlinear coherent state is defined as\cite{Sivakumar98}
\begin{equation}
F(N)a^2|\alpha, F\rangle=\alpha |\alpha, F\rangle, \label{eq:F}
\end{equation}
and the corresponding photon-added two-photon nonlinear coherent state is
\begin{equation}
|\alpha, F, m\rangle=a^{\dagger m}|\alpha, F\rangle
\end{equation}
up to a normailization constant.

Acting the operator $a^2a^{\dagger m}$ on Eq.(\ref{eq:F}) from the left, we obtain
\begin{equation}
F(N-m+2)a^2a^{\dagger m}a^2|\alpha, F\rangle=\alpha (N+1)(N+2)a^{\dagger (m-2)}|\alpha, F\rangle.
\end{equation}

Since
\begin{equation}
a^2a^{\dagger m}a^2=(N+4-m)(N+3-m)a^2a^{\dagger (m-2)},
\end{equation}
we obtain
\begin{equation}
F(N-m+2)(N+4-m)(N+3-m)a^2a^{\dagger (m-2)}|\alpha, F\rangle=\alpha (N+1)(N+2)a^{\dagger (m-2)}|\alpha, F\rangle.
\end{equation}
Let $m-2\rightarrow m$ in the above equation and note that the 
operator $(N+1)(N+2)$ is positive in the whole Fock space, we get
\begin{equation}
F(N-m)(1-\frac{m}{N+2})(1-\frac{m}{N+1})a^2|\alpha, F, m\rangle=\alpha |\alpha, F, m\rangle.
\end{equation}
This shows that the photon-added nonlinear coherent states 
$ |\alpha, F, m\rangle $ are still nonlinear coherent states
with the nonlinear function 
\begin{equation}
F(N-m)(1-\frac{m}{N+2})(1-\frac{m}{N+1}).
\end{equation}

Since the squeezed vaccum state and squeezed first Fock state are two-photon nonlinear coherent state\cite{Sivakumar98},
we conclude that the photon-added squeezed vaccum state and photon-added squeezed first Fock state are 
also two-photon nonlinear coherent states as discussed in Ref.\onlinecite{Liu2000}.  
 We can also introduce the photon-added two-photon nonlinear coherent states with negative 
$m$ and make a similar discussion as one-photon case. We will not explicitly present them here.

\begin{center}
{\bf {\large 6.Conclusions}}
\end{center}

In conclusion, we have studied a special NLCSs, the PANLCSs. From the PANLCS
we introduce a new type of quantum state, the PANLCS with negative values of 
$m$. The states corresponding to the positive and negative values of $m$ are
shown to be the result of nonunitarily deforming the number states $%
|m\rangle $ and $|0\rangle $, respectively. As a example, we study the
sub-Poissonian statistics and squeezing effects in the photon-added
geometric state with negative values of $m$ in detail. The results shows
that photon-added geometric state with negative values of $m$ are always
super-Poissonian and the state can be squeezed in the quadrature $Y.$
We also consider the photon-added two-photon nonlinear coherent states and find a 
similar concusion as one-photon case, i.e, the photon-added two-photon nonlinear coherent states
are still two-photon nonlinear coherent states with certain nonlinear functions.

\vspace{2cm} {\bf Acknowledgment}: The author thanks for the discussions
with Prof. Hong-Chen Fu and help of Prof.Chang-Pu Sun, Shao-Hua Pan, Guo-Zhen Yang. The work is partially supported by the National
Science Foundation of China with grant number:19875008.

\vspace{2cm} {\bf Figure Captions:} \newline
Figure1, Mandel's Q parameter as a function of $\eta $ for different values
of $m.$ \newline
Figure2, Variances of the quadrature operators $X$ and $Y$ as a function of $%
\eta $ for different values of $m.$

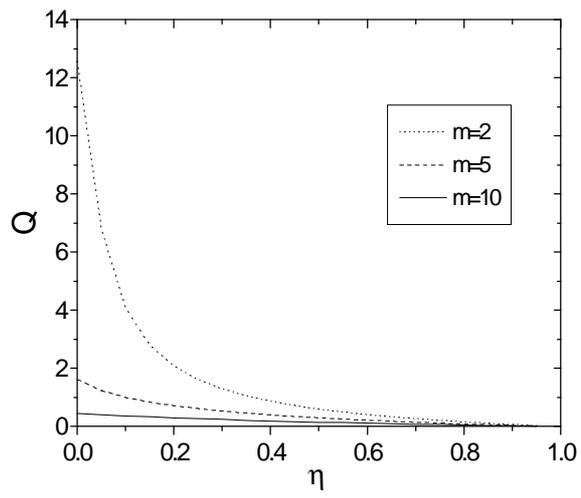
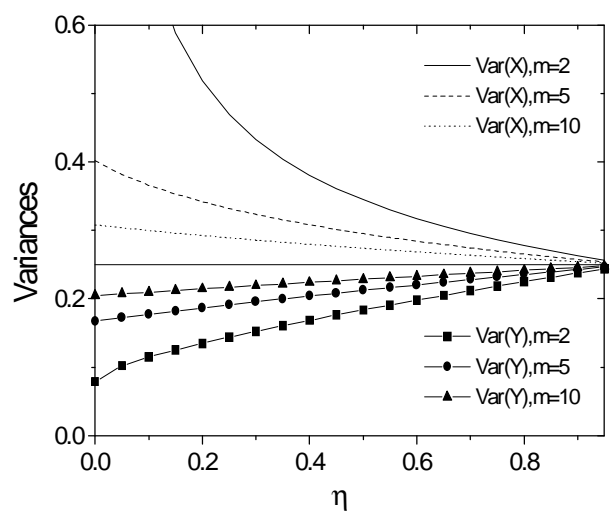
\end{document}